\documentclass[fleqn,usenatbib]{mnras}

\usepackage[margin=5pt]{subfig}
\usepackage{comment}
\usepackage{longtable}
\usepackage{caption}
\usepackage{placeins}
\usepackage{threeparttable}
\usepackage{booktabs}
\usepackage[figuresright]{rotating}
\usepackage{textcase}
\newcommand{\be}{\begin{equation}}
\newcommand{\ee}{\end{equation}}
\newcommand{\bea}{\begin{eqnarray}}
\newcommand{\eea}{\end{eqnarray}}
\newcommand{\hunit}{$\rm{km \ s^{-1} \ Mpc^{-1}}$}
\newcommand{\lcdm}{$\Lambda$CDM}
\usepackage{array}
\makeatletter
\newcommand{\thickhline}{%
    \noalign {\ifnum 0=`}\fi \hrule height 1pt
    \futurelet \reserved@a \@xhline
}
\newcolumntype{"}{@{\hskip\tabcolsep\vrule width 1pt\hskip\tabcolsep}}
\makeatother

\usepackage[T1]{fontenc}
\usepackage{scalerel}
\usepackage{tikz}
\usetikzlibrary{svg.path}

\definecolor{orcidlogocol}{HTML}{A6CE39}
\tikzset{
  orcidlogo/.pic={
    \fill[orcidlogocol] svg{M256,128c0,70.7-57.3,128-128,128C57.3,256,0,198.7,0,128C0,57.3,57.3,0,128,0C198.7,0,256,57.3,256,128z};
    \fill[white] svg{M86.3,186.2H70.9V79.1h15.4v48.4V186.2z}
                 svg{M108.9,79.1h41.6c39.6,0,57,28.3,57,53.6c0,27.5-21.5,53.6-56.8,53.6h-41.8V79.1z M124.3,172.4h24.5c34.9,0,42.9-26.5,42.9-39.7c0-21.5-13.7-39.7-43.7-39.7h-23.7V172.4z}
                 svg{M88.7,56.8c0,5.5-4.5,10.1-10.1,10.1c-5.6,0-10.1-4.6-10.1-10.1c0-5.6,4.5-10.1,10.1-10.1C84.2,46.7,88.7,51.3,88.7,56.8z};
  }
}

\newcommand\orcidicon[1]{\href{https://orcid.org/#1}{\mbox{\scalerel*{
\begin{tikzpicture}[yscale=-1,transform shape]
\pic{orcidlogo};
\end{tikzpicture}
}{|}}}}

\usepackage{hyperref} %<--- Load after everything else

\DeclareRobustCommand{\VAN}[3]{#2}
\let\VANthebibliography\thebibliography
\def\thebibliography{\DeclareRobustCommand{\VAN}[3]{##3}\VANthebibliography}

\usepackage{graphicx}	% Including figure files
\usepackage{amsmath}	% Advanced maths commands
\usepackage{amssymb,bm}	% Extra maths symbols
\usepackage{fnpct}
\usepackage{multirow}
\title[Alleviate $H_0$ tension using CCH and SN Ia data]{Potentialities of Hubble parameter and expansion rate function data to alleviate Hubble tension}

 \author[Yang, Lu, Qian \& Cao]{
 Yingjie Yang$^{\orcidicon{0000-0003-3017-352X}}$,$^{1}$\thanks{E-mail: yyj@smu.edu.cn; (corresponding author)}
 Xuchen Lu$^{2}$, 
 Lei Qian$^{3}$ and
 Shulei Cao$^{\orcidicon{0000-0003-2421-7071}4}$\thanks{E-mail: shulei@phys.ksu.edu; (corresponding author)}
 \\
 $^{1}$Department of Mathematics and Physics, Southern Medical University, Guangzhou 510515, China\\
$^{2}$School of Physics, Huazhong University of Science and Technology, Wuhan 430074, China\\
$^{3}$Guangdong Provincial Key Laboratory of Medical Biomechanics, National Key Discipline of Human Anatomy,\\
School of Basic Medical Sciences, Southern Medical University, Guangzhou 510515, China \\
$^{4}$Department of Physics, Kansas State University, 116 Cardwell Hall, Manhattan, KS 66506, USA\\
  }

\date{Accepted XXX. Received YYY; in original form ZZZ}

\pubyear{2022}

\begin{document}
\label{firstpage}
\pagerange{\pageref{firstpage}--\pageref{lastpage}}
\maketitle

\begin{abstract}
Taking advantage of Gaussian process (GP), we obtain an improved estimate of the Hubble constant, $H_0=70.41\pm1.58$ km s$^{-1}$ Mpc$^{-1}$, using Hubble parameter [$H(z)$] from cosmic chronometers (CCH) and expansion rate function [$E(z)$], extracted from type Ia supernovae, data. We also use CCH data, including the ones with full covariance matrix, and $E(z)$ data to obtain a determination of $H_0=72.34_{-1.92}^{+1.90}$ km s$^{-1}$ Mpc$^{-1}$, which implies that the involvement of full covariance matrix results in higher values and uncertainties of $H_0$. These results are higher than those obtained by directly reconstructing CCH data with GP. In order to estimate the potential of future CCH data, we simulate two sets of $H(z)$ data and use them to constrain $H_0$ by either using GP reconstruction or fitting them with $E(z)$ data. We find that simulated $H(z)$ data alleviate $H_0$ tension by pushing $H_0$ values higher towards $\sim70$ km s$^{-1}$ Mpc$^{-1}$. We also find that joint $H(z)$ + $E(z)$ data favor higher values of $H_0$, which is also confirmed by constraining $H_0$ in the flat concordance model and 2-order Taylor expansion of $H(z)$. In summary, we conclude that more and better-quality CCH data as well as $E(z)$ data can provide a new and useful perspective on resolving $H_0$ tension.
\end{abstract}


\begin{keywords}
cosmological parameters -- dark energy -- cosmology: observations
\end{keywords}
%%%%%%%%%%%%%%%%%%%%%%%%%%%%%%%%%%%%%%%%%%%%%%%%%%

%%%%%%%%%%%%%%%%% BODY OF PAPER %%%%%%%%%%%%%%%%%%

\section{Introduction} \label{sec:intro}

In the past few decades, the fact that our Universe is currently undergoing accelerated expansion is supported by many observations, such as type Ia supernovae (SNe Ia) \citep{Riess:1998cb,Perlmutter:1998np}, cosmic microwave background (CMB) radiation \citep{WMAP:2010qai,Planck:2013pxb,Planck:2018vyg}, large scale structure (LSS) \citep{SDSS:2003eyi}, and baryon acoustic oscillation (BAO) \citep{Beutler:2011hx,Ross:2014qpa,BOSS:2016wmc} measurements. There are many theoretical models proposed to interpret this phenomenon, with the simplest one being the spatially-flat $\Lambda$ cold dark matter ($\Lambda$CDM) model\footnote{In the flat $\Lambda$CDM model, dark energy is a cosmological constant $\Lambda$ that contributes about $70\%$ of the current cosmological energy budget and is the engine for the accelerated expansion.} \citep{peeb84}. Although the flat $\Lambda$CDM model is consistent with most observations, there are potential observational discrepancies (e.g., \citealp{riess_2019}) and theoretical puzzles (e.g., \citealp{Martin}).

Meanwhile as the precision of cosmological observations improve, a tension between measurements of the Hubble constant ($H_0$) determined from Planck CMB anisotropy data \citep{Planck:2018vyg} and those determined from direct local distance ladder measurements \citep{Riess:2021fzl} have emerged. It is unclear whether this tension is caused by some new physics beyond the flat $\Lambda$CDM model, or some systematic effects in either or both of the measurements. In cosmology, the Hubble parameter, $H(z)$ as a function of redshift $z$, is a vital quantity when it comes to the measurements of the cosmological distances such as the luminosity distance ($D_L$) and the angular diameter distance ($D_A$). It is therefore important to measure the Hubble constant, current value of $H(z)$, that can provide definitive information on the scale of the Universe. Consequently, alleviating or resolving the $H_0$ tension becomes crucial. 

There have been many analyses performed trying to explore the values of $H_0$. By reconstructing the $H(z)$ measurements from CCH and from BAO with GP method, \cite{Yu:2017iju} found that $H_0\sim 67\pm4$ \hunit. By using the Gaussian kernel in the GP method, the reconstruction of CCH and SN Ia data from the Pantheon compilation \citep{Pan-STARRS1:2017jku} and the HST CANDELS and CLASH Multi-Cycle Treasury (MCT) programs \citep{Riess:2017lxs} (Pantheon + MCT) derived $H_0=67.06\pm1.68$ \hunit \citep{Gomez-Valent:2018hwc}. By considering variations in total-to-selective extinction of Cepheid flux, hidden structure in the period-luminosity relationship, and intrinsic color distributions of Cepheids, \citep{Follin:2017ljs} determine a value of $H_0=73.3\pm1.7$ \hunit\ that is in agreement with \citep{Riess:2021fzl}. Fitting several physically motivated dark-energy models into simulated calibrators and Pantheon SN Ia data, \citep{Dhawan:2020xmp} find that $H_0$ is not sensitive to pre-selected cosmological models and get the values of $H_0$ to be about 74 \hunit. \citep{Dutta:2019pio} obtain $H_0=70.3_{-1.35}^{+1.36}$ \hunit, consistent with different early and late Universe observations within 2$\sigma$, by fitting spatially flat $\Lambda$CDM model into various low-redshift cosmological data, including SN Ia, BAO, time-delay measurements using strong-lensing, CCH, and growth measurements from large scale structure observations. \citep{DiValentino:2020vnx} get an estimate of $H_0=72.94\pm0.75$ \hunit, which is in 5.9$\sigma$ tension with the \citep{Planck:2018vyg} flat $\Lambda$CDM model value, from 23 $H_0$ measurements based on various sources. Within the Friedmann–Lema\^{i}tre–Robertson–Walker (FLRW) framework, by varying the sound horizon as a free parameter, an upper limit of $H_0\sim71\pm1$ \hunit\ is obtained by \citep{Krishnan:2021dyb}, so in order to meet the local determinations of $H_0\sim73$ \hunit, one plausible solution is to go beyond the FLRW framework. As of now, the Hubble tension is still controversial, so it is necessary to find more clear perspectives on resolving the issue. For this endeavour, we aim to determine the Hubble constant in a model-independent method.

In this paper, we use 31 CCH and 6 expansion rate function, $E(z)$ compressed from Pantheon + MTC SNe Ia, data \citep{Riess:2017lxs} to constrain $H_0$ in a relatively cosmological model-independent way\footnote{\citep{Riess:2017lxs} obtained $E(z)$ data by assuming flat hypersurfaces, so our method here is curvature-dependent.}. By minimizing a $\chi^2$ function defined by $H(z)$ (reconstructed using GP from CCH data) and $E(z)$ (Pantheon + MCT) data, we find that CCH + Pantheon + MCT data provide a value of $H_0=70.41\pm1.58$ \hunit\ that is in slightly better agreement with that of \cite{Riess:2021fzl} than with that of \cite{Planck:2018vyg}. By applying GP reconstruction to CCH data, \cite{Yang:2019fjt} obtained $H_0=67.46\pm4.75$ \hunit. As in \citep{Ma:2010mr}, we simulate two sets of 128 $H(z)$ data using two fiducial models (a forecast of future $H(z)$ observations) and find that GP reconstruction with these simulated $H(z)$ data would provide higher values of $H_0$ that could potentially alleviate the $H_0$ tension. Meanwhile, joint analyses of two simulated $H(z)$ data sets and Pantheon + MCT data provide higher values of $H_0$ and indicate that more $H(z)$ data in the future would help alleviate $H_0$ tension. In addition, we also use CCH and $E(z)$ data to constrain $H_0$ in the flat \lcdm\ model and in a cosmographical approach (Taylor expansion of the Hubble parameter). We find that CCH + Pantheon + MCT data provide estimates of $H_0$ that is also in better agreement with that of \citep{Riess:2021fzl} than with that of \citep{Planck:2018vyg}. Adding Pantheon + MCT data to CCH data seems to result in pushing values of $H_0$ higher and more restrictive.

The paper is organized as follows. In Section \ref{sec:mod} we present the models we used in our analyses.
In Section \ref{sec:data} we briefly introduce the observational data we used. We describe the data analysis methods adopted to constrain $H_0$ in Section \ref{sec:method}. We summarize our results and conclusions in Sections \ref{sec:res} and \ref{sec:conclusion}.

\section{Models}
\label{sec:mod}
In this paper, we use the flat $\Lambda$CDM model and the flat Chevallier-Polarski-Linder (CPL) parametrization \citep{Chevallier:2000qy,Linder:2002et} as fiducial models to simulate two sets of Hubble parameter, $H(z)$, data, where the simulation method is described in detail in Sec. \ref{subsec:sim}. In comparison, we also use flat $\Lambda$CDM and a cosmographical model -- the Taylor expansion of $H(z)$ about redshift $z$ -- to constrain $H_0$. The main features of these models are summarized below.

In the flat $\Lambda$CDM model, the Hubble parameter is
\begin{equation}
H(z)=H_0\sqrt{\Omega_m(1+z)^3+\Omega_{\Lambda}}\equiv H_0E(z),
\end{equation}
where $\Omega_m$ is the current non-relativistic matter density parameter and $\Omega_{\Lambda}=1-\Omega_m$ is the cosmological constant dark energy density parameter.

In the flat CPL parametrization, the Hubble parameter is
\begin{equation}
\resizebox{0.47\textwidth}{!}{%
$H(z)=H_0\sqrt{\Omega_m(1+z)^3+(1-\Omega_m)(1+z)^{3(1+w_0+w_a)}\exp(\frac{-3w_az}{1+z})},$%
}
\end{equation}
where the equation of state parameter is $w(z)=w_0+w_az/(1+z)$ with $w_0$ and $w_a$ being real numbers.

Taylor expansion of the Hubble parameter around present time ($z=0$) is
\begin{equation}
H(z)=H_0+\frac{dH}{dz}\Big|_{z=0}z+\frac{1}{2!}\frac{d^2H}{dz^2}\Big|_{z=0}z^2+\frac{1}{3!}\frac{d^3H}{dz^3}\Big|_{z=0}z^3+\cdots
\end{equation}
where the cosmographical parameters \citep{Capozziello:2011tj}: Hubble parameter $H$, deceleration parameter $q$, jerk parameter $j$, snap parameter $s$, and lerk parameter $l$ are defined as
\begin{equation}
H=\frac{\dot{a}}{a},\quad q=-\frac{\ddot{a}}{aH^2},\quad j=\frac{a^{(3)}}{aH^3},\quad  s=\frac{a^{(4)}}{aH^4},\quad l=\frac{a^{(5)}}{aH^5}.
\end{equation}
In these equations, $a$ is the scale factor, an overdot denotes a time derivative, and $a^{(n)}$ represent the $n$-th time derivative of $a$. Therefore, $H(z)$ can also be expressed as
\begin{equation}
\begin{split}
H(z)=&H_0[1+(1+q_0)z+\frac{1}{2}(j_0-q_0^2)z^2\\
     &+\frac{1}{6}(3q_0^2+3q_0^3-4q_0j_0-3j_0-s_0)z^3\\
     &+\frac{1}{24}(-12q_0^2-24q_0^3-15q_0^4+32q_0j_0+25q_0^2j_0 \\
     &+7q_0s_0+12j_0-4j_0^2+8s_0+l_0)z^4+\mathcal{O}(z^5)   ].
\end{split}
\end{equation}
where the subscript ``0'' indicates that parameters are evaluated at the present epoch. Based on the information provided by \citep{DES:2018rjw} and \citep{Gomez-Valent:2018hwc}, it is reasonable to consider up to fourth-order (excluding first-order) polynomials of Taylor expansion. Since the Taylor expansion of $H(z)$ may break down at high $z$, we only use data with $z\lesssim1.0$ \citep{Gong:2004sd,Zhang:2016urt,OColgain:2021pyh}.

\section{Data}
\label{sec:data}
\cite{Riess:2017lxs} combine the Pantheon SN Ia sample \citep{Pan-STARRS1:2017jku} with 15 SNe Ia at redshift $z>1$ discovered in the CANDELS and CLASH MCT programs \citep{Grogin:2011ua,Koekemoer:2011ub,Postman:2011hg} using WFC3 on the Hubble Space Telescope, and by assuming a flat universe with the curvature energy density parameter $\Omega_k=0$, compress the raw distance measurements into the six expansion rate $E(z)$ measurements in the redshift range $0.07<z<1.5$. The results and the correlation matrix of $E(z)$ are shown in Table \ref{eztable}. Because of the assumption $\Omega_k=0$, the results of $E(z)$ can only be used to constrain spatially-flat cosmological models. Since the last data point $E(z = 1.5)$ is non-Gaussian, the symmetrization of the upper and lower bounds gives $E(1.5)=2.924\pm 0.675$ \citep{Haridasu:2018gqm} or $E(1.5)=2.67\pm 0.675$ \citep{Pinho:2018unz}, and the Gaussian approximation gives $E(1.5)=2.78\pm 0.59$ \citep{Gomez-Valent:2018gvm}. As explained in \cite{Gomez-Valent:2018gvm}, the relative uncertainty of $E(z = 1.5)$ has relatively small impact on the reconstructed functions and treating it as a multivariate Gaussian distribution is more practical, so here we decide to use $E(1.5)=2.78\pm 0.59$. These $E(z)$ data are considered as the compression form of Pantheon + MCT SN Ia data and can significantly reduce the computation time.

\begin{table*}
  \centering
  \caption{$E(z)$ measurements compressed from Pantheon + MCT SNe Ia \protect\citep{Riess:2017lxs}.}
  \setlength{\tabcolsep}{7.3mm}{
  \begin{tabular}{cc|cccccc}
    \hline
     $z$ & $E(z)$ &\multicolumn{6}{c}{Correlation Matrix} \\
    \hline
    0.07 & $0.994\pm 0.023$ & 1.00 &  &  &  &  &  \\
    0.2  & $1.113\pm 0.020$ & 0.40 & 1.00 &  &  &  & \\
    0.35 & $1.122\pm 0.037$ & 0.52 & -0.13 & 1.00 &  &  & \\
    0.55 & $1.369\pm 0.063$ & 0.35 & 0.35 & -0.18 & 1.00 &  & \\
    0.9  & $1.54\pm 0.12$ & 0.02 & -0.08 & 0.19 & -0.41 & 1.00 & \\
    1.5  & $2.69^{+0.86}_{-0.52}$ & 0.00 & -0.06 & -0.05 & 0.16 & -0.21 & 1.00\\
    \hline
  \end{tabular}}
  \label{eztable}
\end{table*}

The 31 CCH data\footnote{For cosmological analyses using $H(z)$ data, see e.g. \citep{Caoetal_2018a,Caoetal_2018b,Caoetal_2018c,Ryan_1,CaoRyanRatra2020,Koksbang:2021qqc,CaoRyanRatra2021,CaoZWZ2021,Caoetal_2021,CaoRyanRatra2022,CaoKhadkaRatra2022,Cao:2022wlg,CaoRatra2022,Caoetal2022}.} \citep{Yang:2019fjt} that reach to $z\sim 2$ are listed in Table \ref{hztable} and are cosmological model-independent. In this paper we use both $E(z)$ and CCH data to perform our analyses. The constraints on $H_0$ obtained from these data are free of local $H_0$ measurements and the early-Universe observations like CMB, so can be used as comparisons. In addition, we also use the suggested full covariance matrix of 15 CCH data \citep{Moresco:2020fbm,Moresco:2022phi} to perform some of our analyses.

\begin{table*}
\centering
\caption{The 31 CCH data. The unit for $H(z)$ is \hunit.}
\setlength{\tabcolsep}{4mm}{
\begin{tabular}{ccccccccc}
\toprule
$z$ & $H(z)$ & $\sigma_{H}$& Ref. & &$z$ & $H(z)$ & $\sigma_{H}$& Ref. \\
\midrule
 0.07 & 69.0 & 19.6 & \cite{Zhang:2012mp} & &0.4783 & 80.9 & 9.0  & \cite{Moresco:2016mzx}\\
 0.09 & 69.0 & 12.0 & \cite{Simon:2004tf} & & 0.48 & 97.0 & 62.0 & \cite{Stern:2009ep}  \\
 0.12 & 68.6 & 26.2 & \cite{Zhang:2012mp}  & & 0.593 & 104.0 & 13.0 & \cite{Moresco:2012jh} \\
 0.17 & 83.0 & 8.0 & \cite{Simon:2004tf}  & & 0.68 & 92.0 & 8.0 & \cite{Moresco:2012jh} \\
 0.179 & 75.0 & 4.0 & \cite{Moresco:2012jh}  & & 0.781 & 105.0 & 12.0  & \cite{Moresco:2012jh} \\
 0.199 & 75.0 & 5.0 & \cite{Moresco:2012jh}  & & 0.875 & 125.0 & 17.0 & \cite{Moresco:2012jh} \\
 0.2 & 72.9 & 29.6 & \cite{Zhang:2012mp}  & & 0.88 & 90.0 & 40.0 & \cite{Stern:2009ep} \\
 0.27 & 77.0 & 14.0 & \cite{Simon:2004tf}  & & 0.9 & 117.0 & 23.0 & \cite{Simon:2004tf} \\
 0.28 & 88.8 & 36.6 & \cite{Zhang:2012mp}  & & 1.037 & 154.0 & 20.0 & \cite{Moresco:2012jh} \\
 0.352 & 83.0 & 14.0 & \cite{Moresco:2012jh}  & & 1.3 & 168.0 & 17.0 & \cite{Simon:2004tf} \\
 0.3802 & 83.0 & 13.5 & \cite{Moresco:2016mzx} & & 1.363 & 160.0 & 33.6 & \cite{Moresco:2015cya} \\
 0.4 & 95.0 & 17.0 & \cite{Simon:2004tf}  & & 1.43 & 177.0 & 18.0 & \cite{Simon:2004tf} \\
 0.4004 & 77.0 & 10.2 & \cite{Moresco:2016mzx}  & & 1.53 & 140.0 & 14.0 & \cite{Simon:2004tf} \\
 0.4247 & 87.1 & 11.2  & \cite{Moresco:2016mzx} & & 1.75 & 202.0 & 40.0 & \cite{Simon:2004tf} \\
 0.4497 & 92.8 & 12.9  & \cite{Moresco:2016mzx} & & 1.965 & 186.5 & 50.4  & \cite{Moresco:2015cya} \\
 0.47 & 89.0 & 49.6  & \cite{Ratsimbazafy:2017vga} & & & & & \\
 \bottomrule
\end{tabular}%
}
\label{hztable}
\end{table*}

\section{Method}
\label{sec:method}
\subsection{Gaussian process}
\label{subsec:GP}
The Gaussian process (GP) method is a powerful model-independent tool to reconstruct a function and its derivatives from discrete data points, which has been widely used in cosmology to probe the property of cosmic acceleration \citep{Clarkson:2010bm,Shafieloo:2012ht,Holsclaw:2010nb,Holsclaw:2010sk,Holsclaw:2011wi,Seikel:2012uu,Nair:2013sna}, 
test the concordance model \citep{Seikel:2012cs,Bilicki:2012ub,Vitenti:2015aaa,Yahya:2013xma,Yennapureddy:2017vvb,Yennapureddy:2018rdz}, and reconstruct $H_0$ \citep{Busti:2014dua,Sahni:2014ooa,Verde:2014qea,Wang:2016iij,Zhang:2016tto,Yu:2017iju,Yennapureddy:2017vvb,Melia:2018tzi,Gomez-Valent:2018hwc,Jesus:2019nnk}.

The GP reconstruction is determined by a mean function with Gaussian error bars, where the values of the function at different redshifts $z$ and $z'$ are correlated through a covariance function $k(z,z')$. Here we use the squared-exponential covariance function defined by
\begin{equation}
\label{covf}
k(z,z')=\sigma_f^2\exp\left(-\frac{(z-z')^2}{2l^2}\right),
\end{equation}
where $\sigma_f$ and $l$ are hyperparameters, which are related to the strength of the correlation of the function's value and to the coherence length of the correlation in the input space, respectively. Here we use the public available \textsc{python} package GaPP \citep{Seikel:2012uu} to perform the GP reconstruction.\footnote{The more detailed descriptions of the GP method can be found in section 2 of \citep{Seikel:2012uu}.}

\subsection{Simulation of $H(z)$ data}
\label{subsec:sim}
\begin{figure*}
\centering
\includegraphics[width=0.8\linewidth]{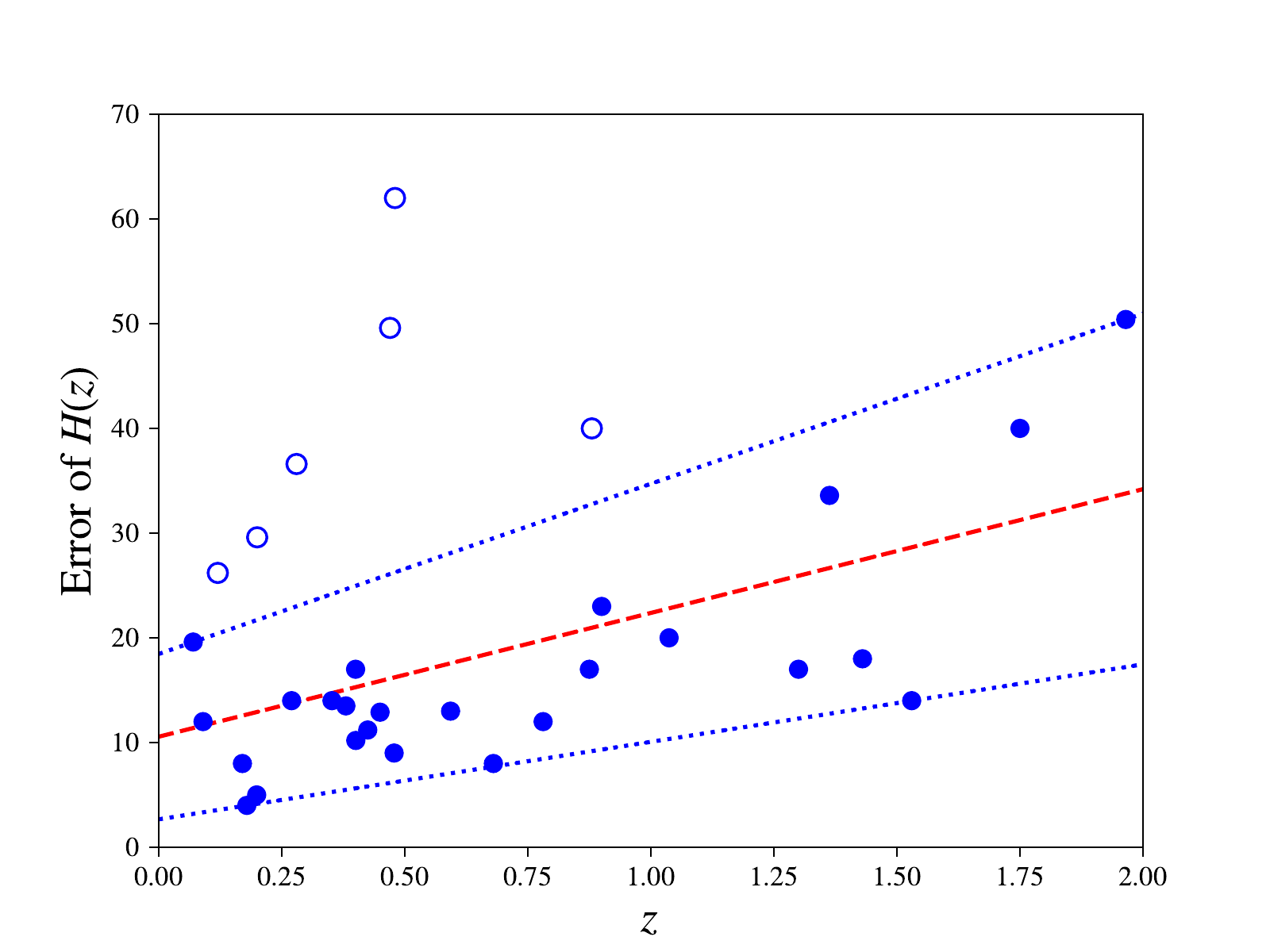}
\caption{The errors of $H(z)$ in 31 CCH data. The solid dots and circles represent non-outliers and outliers, respectively. The dotted lines and the dashed line correspond to Error Model 1 of the bounds $\sigma_+=16.25z+18.46$ and $\sigma_-=7.40z+2.67$, and the mean uncertainty $\sigma_0=11.82z+10.56$, respectively.}
\label{errhz}
\end{figure*}

Following the method used in \cite{Ma:2010mr}, we update the errors of current cosmic chronometers data and identify 6 data points as outliers to be excluded. We then use the remaining CCH data to estimate the errors of the simulated data. Assuming that the errors of $H(z)$ increase linearly with respect to $z$, we find that the upper and lower bounds for the uncertainties $\sigma(z)$ are $\sigma_+(z)=16.25z+18.46$ and $\sigma_-(z)=7.40z+2.67$, respectively, as shown in figure \ref{errhz}. As an estimate of the mean uncertainty for future observations, the midline of the errors is $\sigma_0=11.82z+10.56$. A value of simulated $H(z)$ is generated by $H_{\rm sim}(z)=H_{\rm fid}(z)+\mathcal{N}(0,\tilde{\sigma}(z))$, where $H_{\rm fid}(z)$ is the value of $H(z)$ computed from the fiducial model, and $\mathcal{N}(0,\tilde{\sigma}(z))$ is a random number generated from a Gaussian distribution with mean zero and variance $\tilde{\sigma}(z)$. The variance $\tilde{\sigma}(z)$ is generated randomly from the Gaussian distribution $\mathcal{N}(\sigma_0(z),\epsilon(z))$, where $\epsilon(z)=(\sigma_+(z)-\sigma_-(z))/4$ is chosen to ensure that the probability of $\tilde{\sigma}(z)$ falling in the regions between $\sigma_+(z)$ and $\sigma_-(z)$ is $95.4\%$.

\begin{figure*}
\centering
\includegraphics[width=0.8\linewidth]{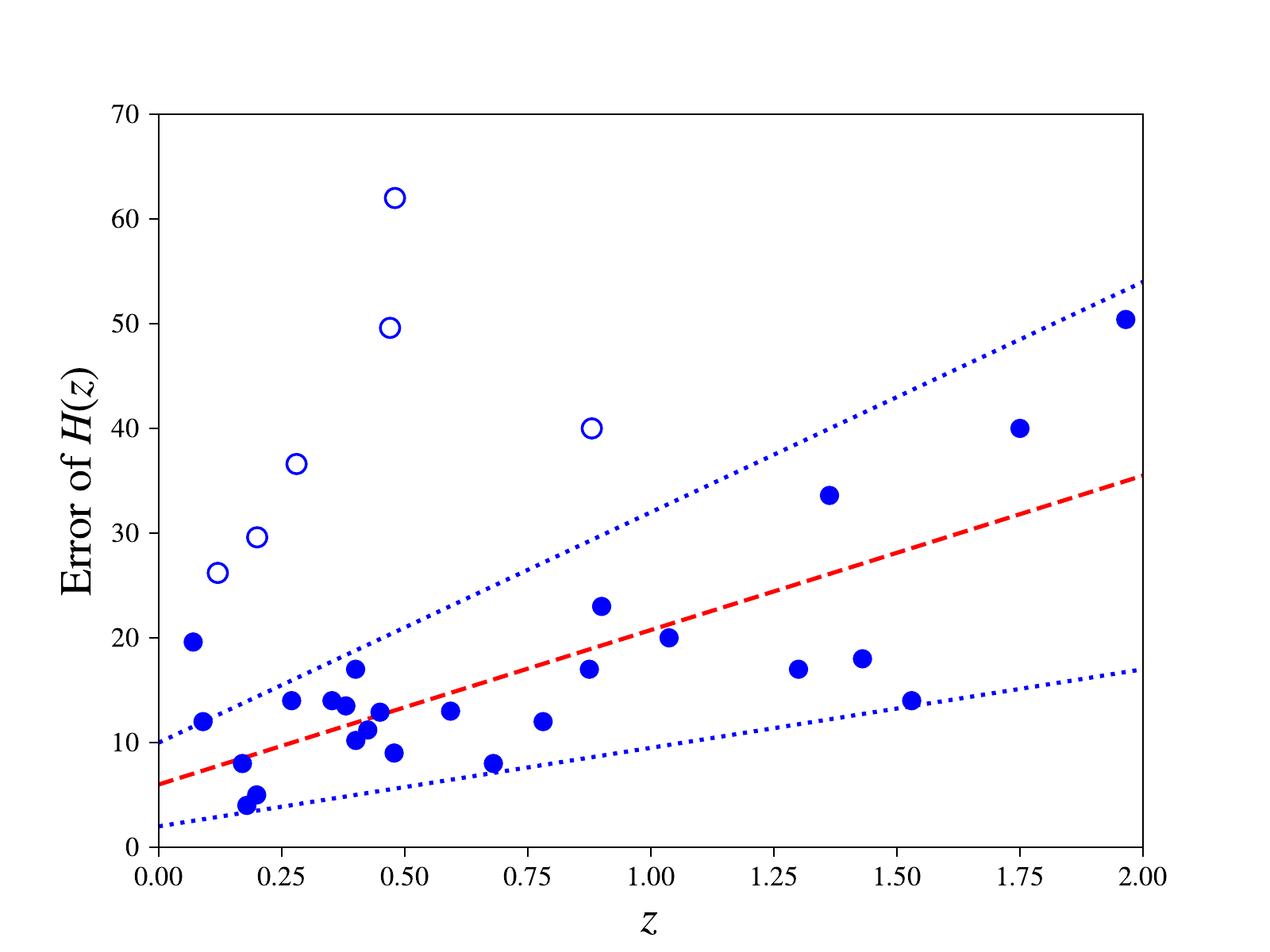}
\caption{Same as Fig. \ref{errhz}, but for Error Model 2 of $\sigma_+=22.00z+10.00$, $\sigma_-=7.50z+2.00$, and $\sigma_0=14.75z+6.00$.}
\label{errhzlinear1}
\end{figure*}

\begin{figure*}
\centering
\includegraphics[width=0.8\linewidth]{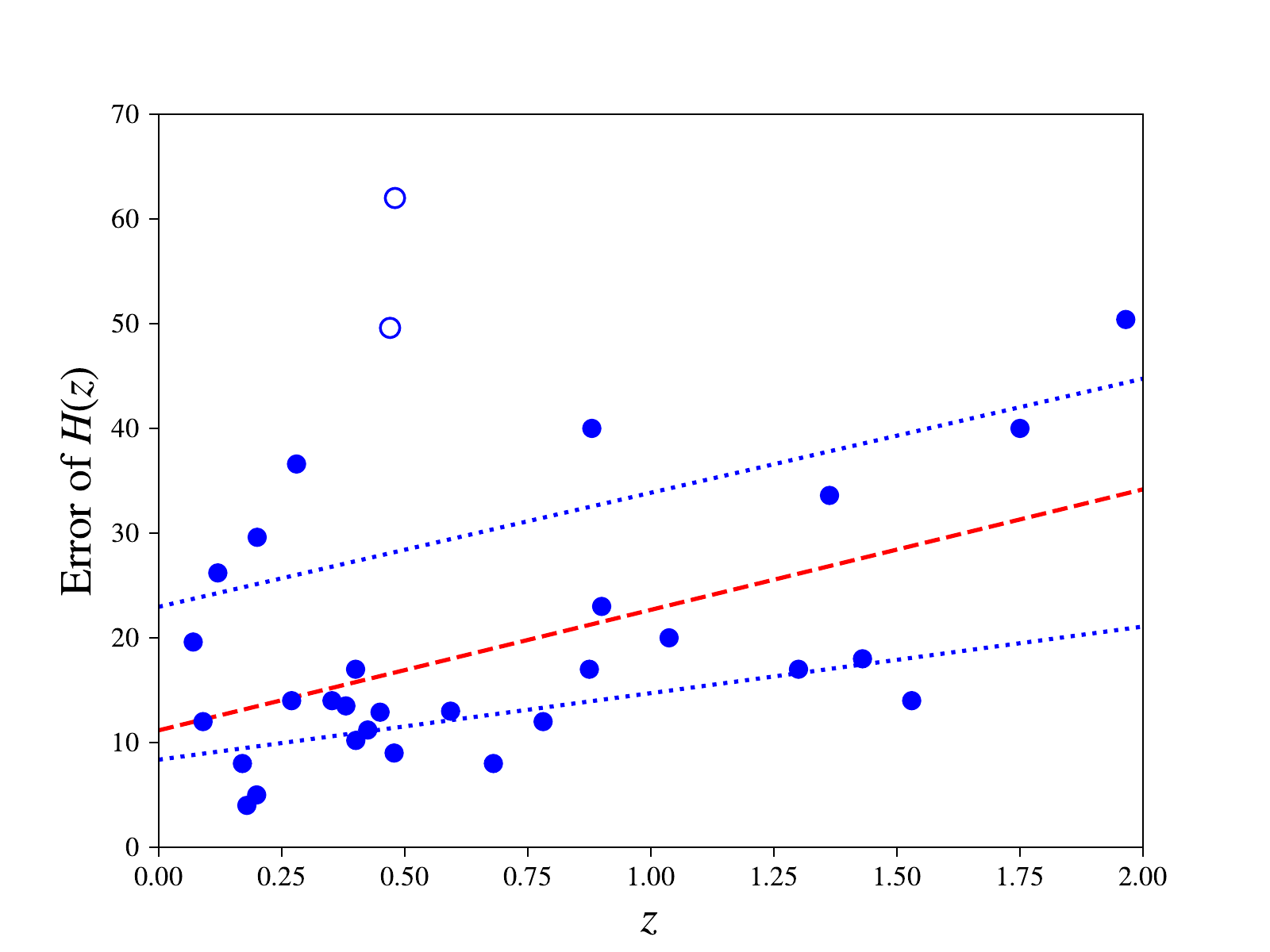}
\caption{Same as Fig. \ref{errhz}, but for Error Model 3 of $\sigma_+=10.89z+22.96$, $\sigma_-=6.35z+8.37$, and $\sigma_0=11.51z+11.17$ with only two outliers.}
\label{errhzlinear2}
\end{figure*}

\begin{figure*}
\centering
\includegraphics[width=0.8\linewidth]{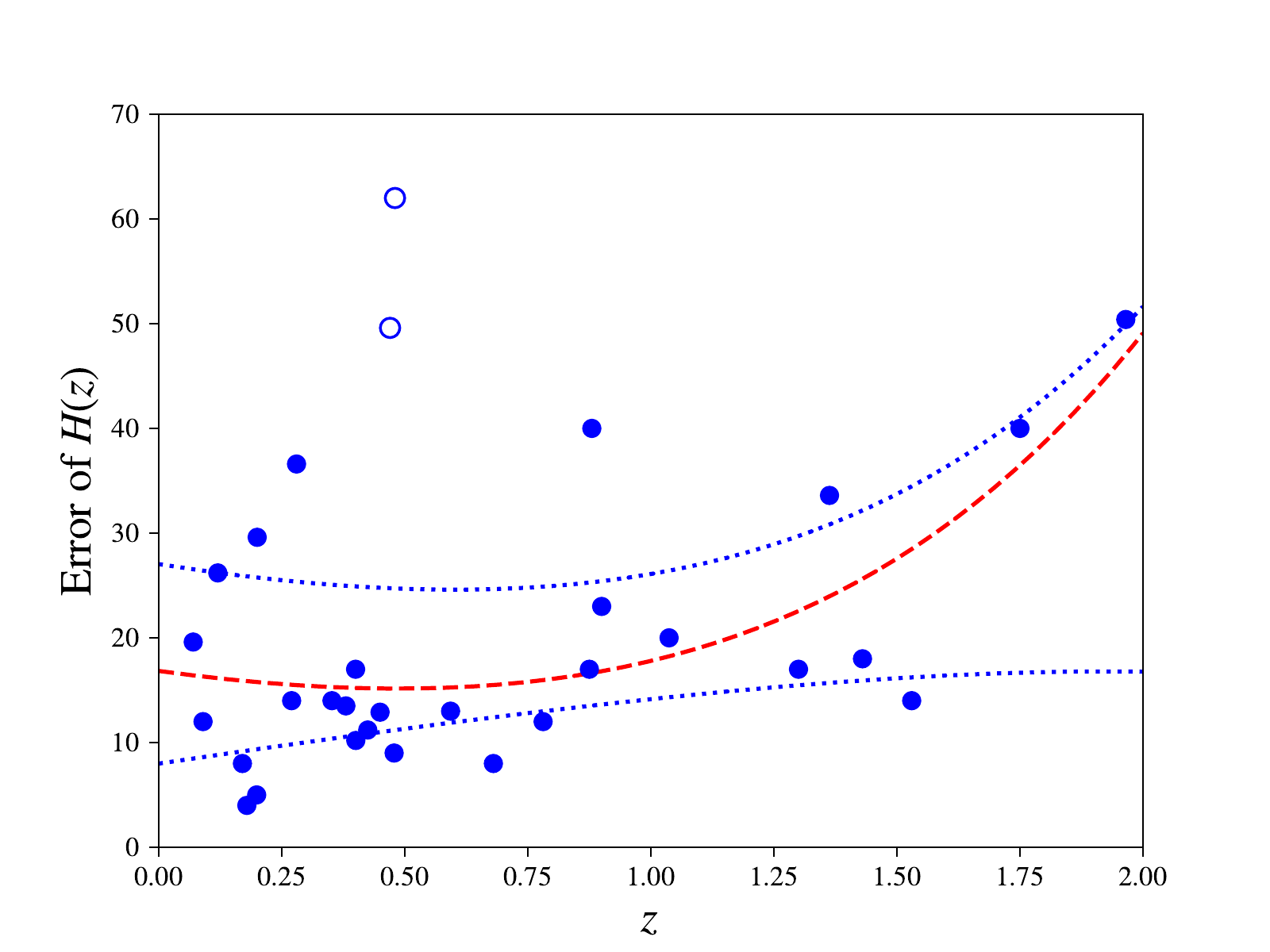}
\caption{Same as Fig. \ref{errhz}, but for Error Model 4 of $\sigma_+=8.97e^z-16.36z+18.07$, $\sigma_-=-1.20e^z+8.21z+9.19$, and $\sigma_0=10.28e^z-16.70z+6.57$.}
\label{errhznonlinear}
\end{figure*}

In order to test the robustness of this linear error model (Error Model 1), in comparison, we choose two other linear models and one non-linear model to simulate $H(z)$ data. In the first linear model (Error Model 2), as shown in Fig. \ref{errhzlinear1}, the red dashed line is $\sigma_0=14.75z+6.00$, obtained by applying linear regression method to the errors of 25 solid $H(z)$ data, while the two blue dotted lines are $\sigma_+=22.00z+10.00$ and $\sigma_-=7.50z+2.00$, selected symmetrically around $\sigma_0$ to ensure that most data points are in between.

In the second linear model (Error Model 3), as shown in Fig. \ref{errhzlinear2}, the red dashed line is $\sigma_0=11.51z+11.17$, obtained by applying linear regression method to the errors of 29 solid $H(z)$ data. Then we apply linear regression method to data points above and below $\sigma_0$ to derive $\sigma_+=10.89z+22.96$ and $\sigma_-=6.35z+8.37$, respectively.

In the non-linear model (Error Model 4), as shown in Fig. \ref{errhznonlinear}, $\sigma_0=10.28e^z-16.70z+6.57$ is best fitted by a randomly chosen form of exponential and linear function from the 29 solid $H(z)$ data. Best fittings of $\sigma_+=8.97e^z-16.36z+18.07$ and $\sigma_-=-1.20e^z+8.21z+9.19$ are derived from the data points above and below $\sigma_0$, respectively.

\begin{figure*}
\centering
\includegraphics[width=0.8\linewidth]{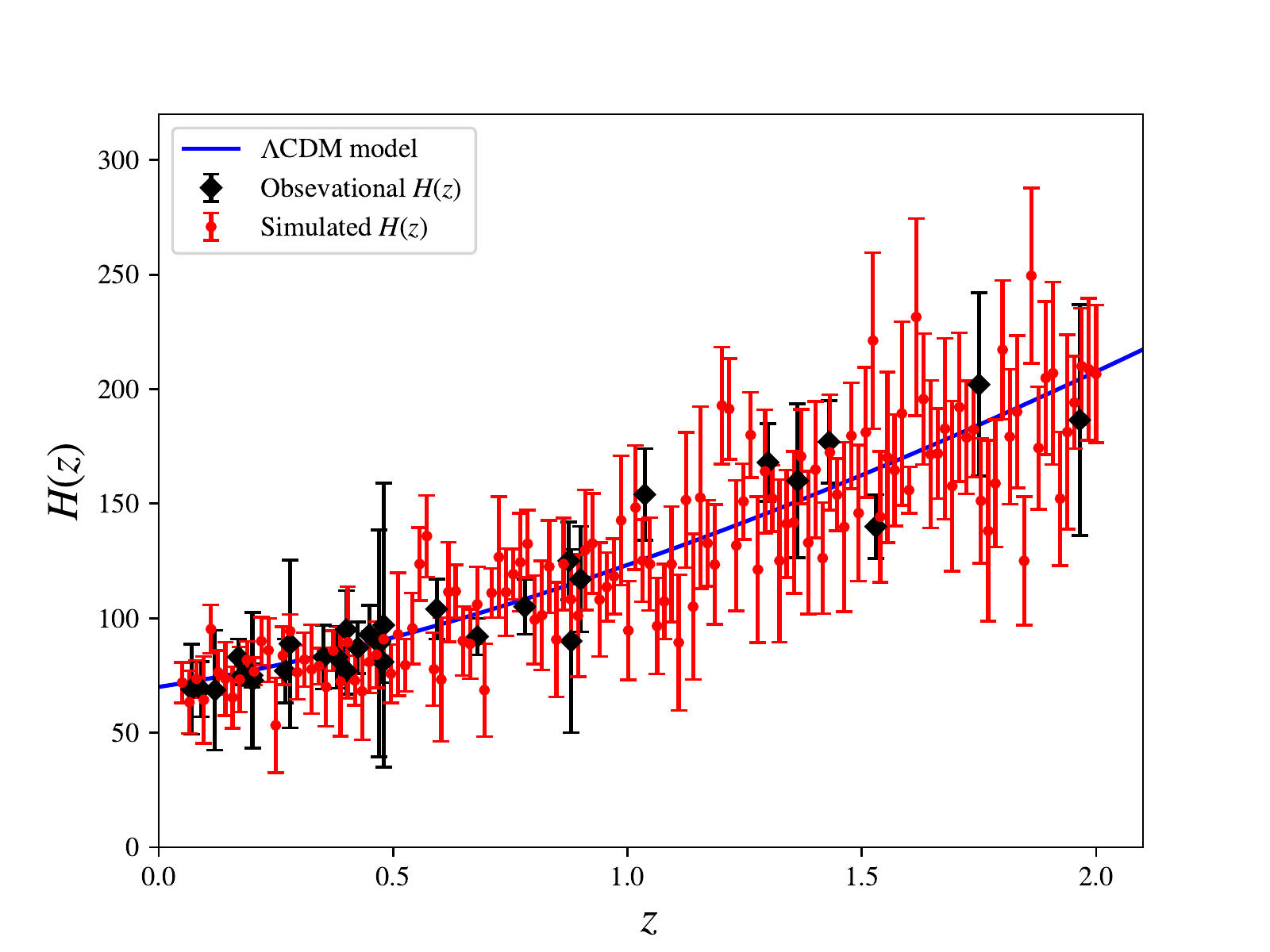}
\caption{The simulated data set of $H(z)$ in the redshift $0.05<z<2.0$. The red points with error bars represent the simulated $H(z)$ data, the diamond points with error bars represent the cosmic chronometers data, and the blue line represent the fiducial model, i.e. flat \lcdm\ with $\Omega_m=0.3$ and $H_0=70$ \hunit.}
\label{mockhz}
\end{figure*}

\begin{figure*}
\centering
\includegraphics[width=0.8\linewidth]{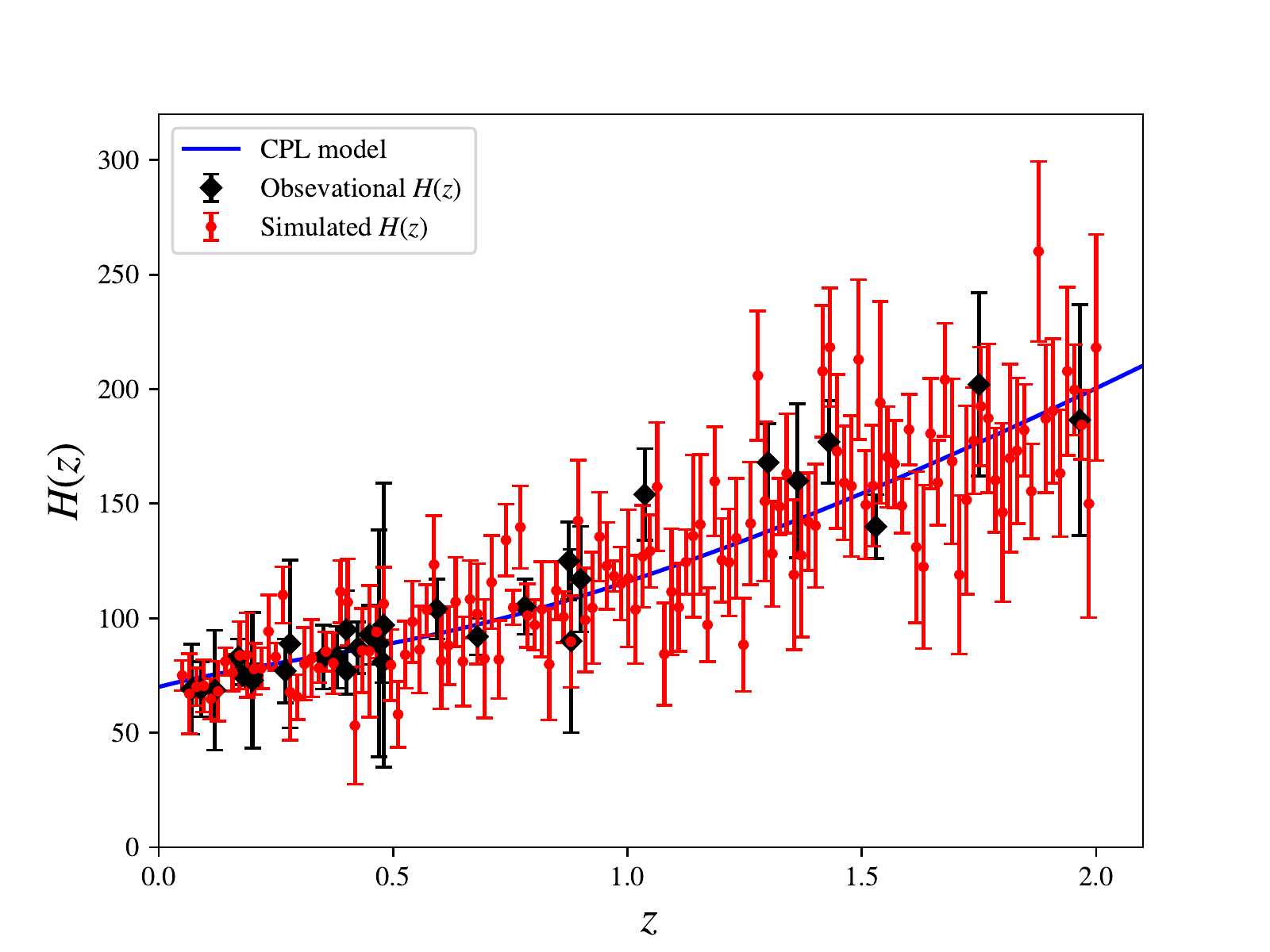}
\caption{Same as Fig. \ref{mockhz} but with the flat CPL parametrization as fiducial model, where $\Omega_m=0.3$, $H_0=70$ \hunit, $w_0=-0.705$, and $w_a=-2.286$.}
\label{mockCPL}
\end{figure*}

We use flat $\Lambda$CDM with $\Omega_m=0.3$ and $H_0=70$ \hunit\ as a fiducial model. The 128 simulated data within redshift range of $0.05<z<2.0$ are shown in Fig. \ref{mockhz}. In comparison, we also consider flat CPL parametrization with $w_0=-0.705$ and $w_a=-2.286$ \citep{Hu:2015ksa,Zhang:2016tto} as a fiducial model. The resulting 128 simulated data points within redshift range of $0.05<z<2.0$ are shown in Fig. \ref{mockCPL}. In order to investigate the influence of $H_0$ prior on the reconstruction, we also use $H_0=67.4$ and 73.2 \hunit\ as priors in the two fiducial models for simulations.

\subsection{$\chi^2$ minimization}
\label{subsec:chi2min}
In order to take full advantage of the SNe Ia information, we combine the 31 CCH data with the 6 $E(z)$ data (dubbed ``Pantheon + MCT'') to constrain $H_0$. We apply GP method to reconstruct $H(z)$ from the 31 CCH data and obtain the smoothed $H_{\rm \textsc{gp}}(z)$ function. The GP reconstructed expansion rate is therefore
\begin{equation}
\label{Egp}
E_{\rm \textsc{gp}}(z)=\frac{H_{\rm \textsc{gp}}(z)}{H_0},
\end{equation}
and by treating $H_0$ as a free parameter, the error of $E_{\rm \textsc{gp}}(z)$ is
\begin{equation}
\label{errEgp}
\sigma_{E_{\rm \textsc{gp}}}=\frac{\sigma_{H_{\rm \textsc{gp}}}}{H_0},
\end{equation}
where $\sigma_{H_{\rm \textsc{gp}}}$ at a given $z$ is the corresponding error of $H_{\rm \textsc{gp}}(z)$.

We treat the reconstructed $E_{\rm \textsc{gp}}(z)$ as ``theoretical'' predictions of $E(z)$ and then compare them with the measured values of $E_{\rm obs}(z)$ compressed from Pantheon + MCT SNe Ia by using the $\chi^2$ function
\begin{equation}\label{chi2}
\chi^2=[E_{\rm obs}(z_i)-E_{\rm \textsc{gp}}(z_i)]^T\mathbfss{C}^{-1}[E_{\rm obs}(z_i)-E_{\rm \textsc{gp}}(z_i)],
\end{equation}
where $\mathbfss{C}=\mathbfss{C}_{E}+\mathrm{diag}(\sigma_{E_{\mathrm{\textsc{gp}}}}^2)$ is the total covariance matrix and $\mathbfss{C}_{E}$ is the covariance matrix of $E_{\rm obs}$. We use Markov chain Monte Carlo (MCMC) package \textsc{emcee} \citep{emcee} to constrain $H_0$ by minimizing the $\chi^2$ function in equation \eqref{chi2}. It is worth noting that this approach is cosmological model-independent, except that the observed $E(z)$ data are somewhat curvature-free. Previous studies have used similar approach to measure the cosmic curvature \citep{Wei:2016xti,Wei:2018cov,Li:2016wjm,Yang:2020bpv}.

To better understand the effect of CCH and Pantheon + MCT on the constraints of $H_0$, we also use the flat $\Lambda$CDM model to fit the 31 CCH and 6 $E(z)$ data.
The $\chi^2$ function of $H(z)$ is
\begin{equation}
\chi_H^2=\sum_i\frac{[H_{\rm obs}(z_i)-H_{\rm th}(z_i)]^2}{\sigma_{H,i}^2},
\end{equation}
with $\sigma_{H,i}$ being the uncertainty of $H_{\rm obs}(z_i)$ from CCH data, and the $\chi^2$ function of $E(z)$ is 
\begin{equation}
  \chi_E^2=[E_{\rm obs}(z_i)-E_{\rm {th}}(z_i)]^T\mathbfss{C}_{E}^{-1}[E_{\rm obs}(z_i)-E_{\rm {th}}(z_i)].
\end{equation}
For joint analysis, the total $\chi^2$ is given by
\begin{equation}
\chi_{\rm tot}^2=\chi_H^2+\chi_E^2
\end{equation}

We compare our $H_0$ results with the results from \cite{Planck:2018vyg} ($H_0=67.4\pm0.5$ \hunit) and \cite{Riess:2021fzl} (R21, $H_0=73.2\pm1.3$ \hunit) by computing their differences in units of $\sigma$ in quadrature sum.

\section{Results}
\label{sec:res}

The constraints on $H_0$ from GP reconstruction and relatively cosmological model-independent $\chi^2$ minimization are listed in column 4 of Table \ref{sumh0}. \cite{Busti:2014dua} obtained a GP reconstruction value of $H_0=64.9\pm4.2$ \hunit\ from 19 CCH data. \cite{Yang:2019fjt} obtained a GP reconstruction value of $H_0 = 67.46 \pm 4.75$ \hunit\ from 31 CCH data. By taking into account the full covariance matrix of 15 CCH data provided by Michele Moresco \citep{Moresco:2020fbm,Moresco:2022phi} to form a so-called $H(z)$mat data set, here we obtain a GP reconstruction value of $H_0=67.06\pm4.66$ \hunit, which has mildly smaller central value and uncertainty than the one from \cite{Yang:2019fjt}. We find that as the number of $H(z)$ data increases, the GP reconstructed $H_0$ increases. Motivated by this trend, we use two simulated $H(z)$ data sets from two different fiducial models to perform the GP reconstruction. We find that GP reconstructions from 128 simulated $H(z)$ data with flat \lcdm\ and flat CPL as fiducial models give $H_0 = 71.10 \pm 3.58$ \hunit\ and $H_0 = 71.18\pm3.16$ \hunit, respectively. This indicates that in the future more $H(z)$ data could have the ability to alleviate $H_0$ tension.

\begin{figure*}
\centering
\includegraphics[width=0.8\linewidth]{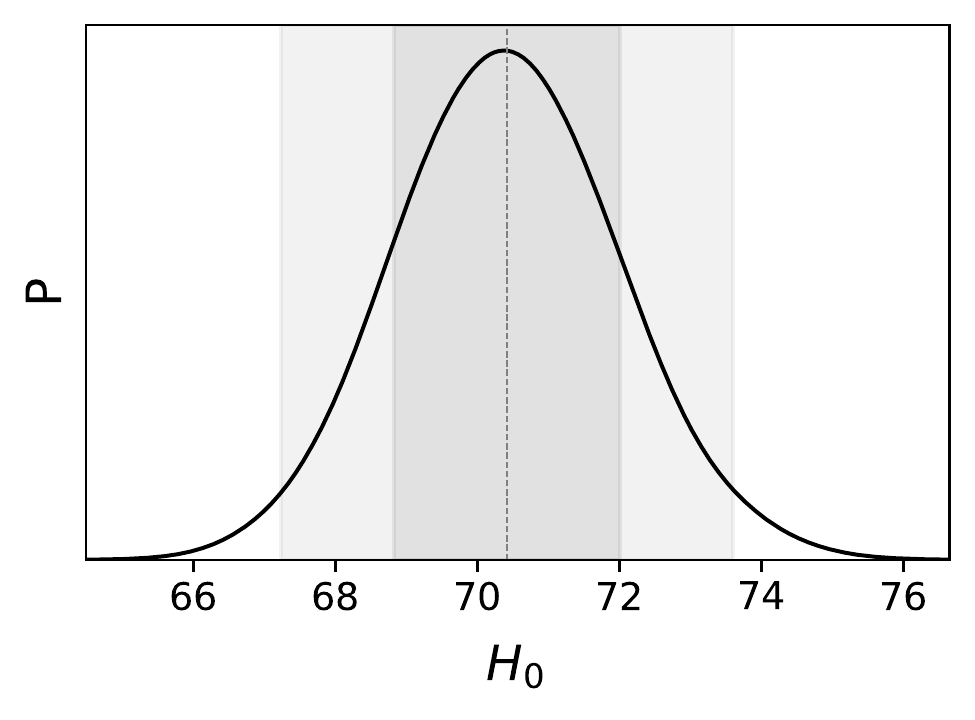}
\caption{The posterior one-dimensional probability distribution of $H_0$ from CCH + Patheon + MCT data by minimizing the $\chi^2$ function \eqref{chi2}.}
\label{h0mc}
\end{figure*}

\begin{figure*}
\centering
\includegraphics[width=0.8\linewidth]{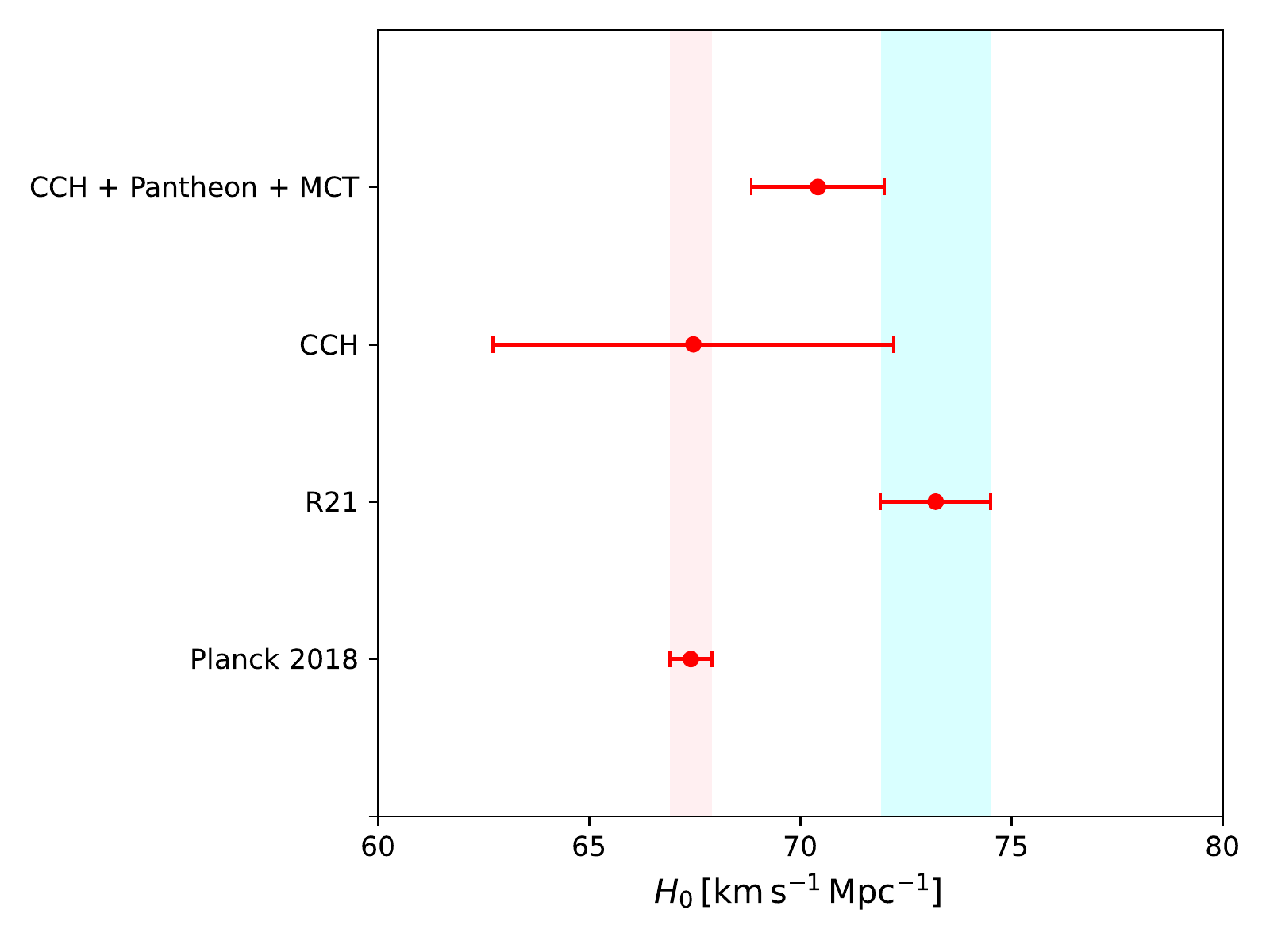}
\caption{$68\%$ constraints of the Hubble constant $H_0$ from different methods, where CCH and CCH + Pantheon + MCT correspond to the Gaussian process reconstruction from the corresponding data sets. The light pink and cyan vertical bands represent the flat \lcdm\ Planck TT,TE,EE+lowE+lensing $H_0$ value \protect\citep{Planck:2018vyg} ($H_0=67.4\pm0.5$ \hunit) and the local $H_0$ value from SH0ES team \protect\citep{Riess:2021fzl} (R21, $H_0=73.2\pm1.3$ \hunit). }
\label{h0}
\end{figure*}

By taking advantage of the GP method and $E(z)$ (Pantheon + MCT) data, we use GP reconstructed $H(z)$ and $E(z)$ data to minimize the $\chi^2$ function \eqref{chi2} and obtain the constraints of Hubble constant. From Table \ref{sumh0}, we can see that CCH + Pantheon + MCT data provide $H_0=70.41\pm1.58$ \hunit, flat \lcdm\ simulated $H(z)$ + Pantheon + MCT data provide $H_0=72.11\pm1.43$ \hunit, and flat CPL simulated $H(z)$ + Pantheon + MCT data provide $H_0=71.34\pm1.39$ \hunit, which are in slightly better agreement with $H_0$ value of R21 than with that of Planck. The posterior one-dimensional probability distribution of $H_0$ from CCH + Patheon + MCT data is shown in Fig. \ref{h0mc} and the values of $H_0$ obtained from different methods are shown in Fig. \ref{h0}. Compared with direct GP reconstruction of $H_0$ from CCH or simulated data, the addition of $E(z)$ data makes the constraints on $H_0$ higher and more restrictive. It is clear that here $E(z)$ data play a dominant role and prefer higher values of $H_0$. Therefore, although joint analyses of simulated $H(z)$ data and $E(z)$ data are in tension with Planck result to $\sim 3\sigma$, the significance of this tension is not definitive due to the lack of the actual $H(z)$ data\footnote{Simulated $H(z)$ data depend on the fiducial models so can only be used as a reference and for qualitative analyses purpose.} and curvature-free of $E(z)$ data.

In order to penalize preferred $H_0$ values for deviating from $E(z=0)=1$, we add an additional term
\begin{equation*}
    \frac{\left[E(z=0)-E_{\rm GP}(z=0)\right]^2}{\sigma^2_{E_{\rm GP}(z=0)}}\equiv\frac{\left[1-E_{\rm GP}(z=0)\right]^2}{\sigma^2_{E_{\rm GP}(z=0)}}
\end{equation*}
to our original $\chi^2$ function \eqref{chi2}. We find that adding this term results in lower values of $H_0$. Specifically, $H(z) + E(z) + E_0$ provides $H_0=70.13\pm1.49$ \hunit, while $H(z)\mathrm{mat} + E(z)$ and $H(z)\mathrm{mat} + E(z) + E_0$ provide $H_0=72.34^{+1.90}_{-1.92}$ and $H_0=71.56\pm1.79$, respectively. Including 15 CCH data with full covariance matrix results in slightly higher values and uncertainties of $H_0$.

\begin{figure*}
\centering
\includegraphics[width=0.8\linewidth]{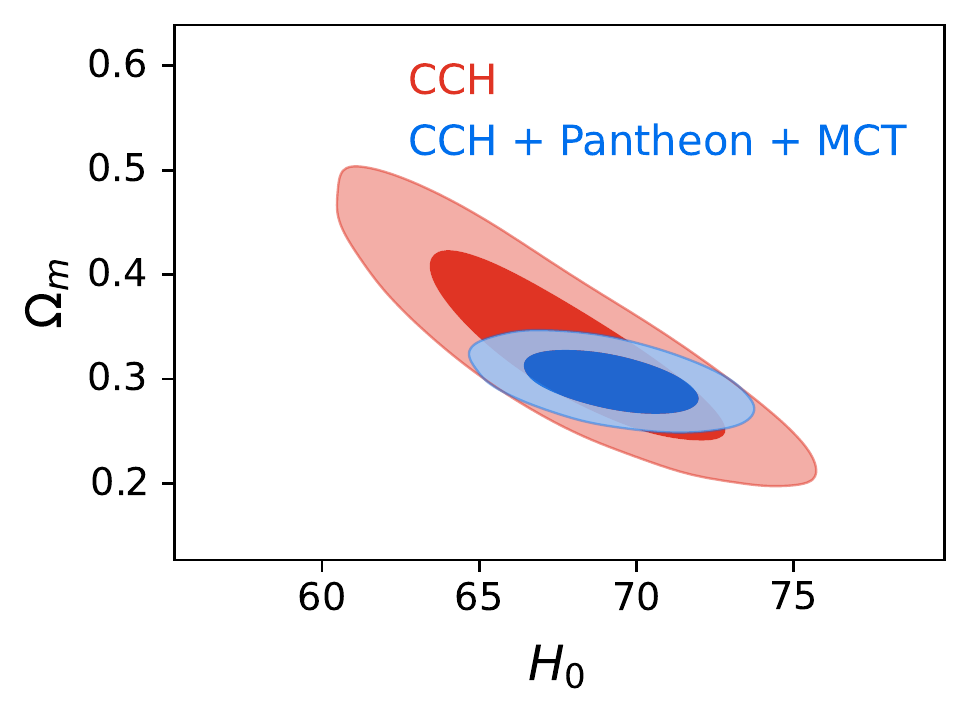}
\caption{The $1\sigma$ and $2\sigma$ confidence regions for the flat $\Lambda$CDM model with constraints from CCH data and CCH + Panthen + MCT data in red and blue, respectively. }
\label{LCDMmc}
\end{figure*}

We also use CCH and CCH + Pantheon + MCT data to constrain cosmological parameters in flat \lcdm\ and cosmographical parameters in Taylor expansion of $H(z)$. The constraints on $H_0$ are listed in column 3 of Table \ref{modeltable}. The flat \lcdm\ $\Omega_m-H_0$ contours are shown in Fig. \ref{LCDMmc}. Cosmological parameter constraints are $\{H_0, \Omega_m\}=\{67.77\pm3.13, 0.321\pm0.063\}$ from CCH data and $\{H_0, \Omega_m\}=\{69.19\pm1.84, 0.297\pm0.020\}$ from CCH + Pantheon + MCT data in the flat \lcdm\ model. $H(z)$mat and $H(z)$mat + Pantheon + MCT data provide cosmological constraints of $\{H_0, \Omega_m\}=\{68.95\pm4.12, 0.324_{-0.072}^{+0.049}\}$ and $\{H_0, \Omega_m\}=\{70.19\pm2.61, 0.297\pm0.020\}$, respectively, which again implies that including 15 CCH data with full covariance matrix indeed results in slightly higher values and uncertainties of $H_0$. This confirms the higher-value-preference of $H_0$ by Pantheon + MCT SN Ia data. In the 2-order $H(z)$ Taylor expansion case, CCH data favor a lower value of $H_0=66.50\pm3.92$ \hunit\ and CCH + Pantheon + MCT data favor a slightly higher and more restrictive value of $H_0=68.45\pm1.90$ \hunit\ that is in better agreement with Planck result than with R21 result. While in the 3-order and 4-order $H(z)$ Taylor expansion cases, CCH data provide very loose constraints of $H_0=70.63\pm8.26$ \hunit\ and $H_0=68.00^{+8.00}_{-10.00}$ \hunit, respectively, and CCH + Pantheon + MCT data provide values of $H_0=68.62\pm1.96$ \hunit\ and $H_0=68.75\pm1.97$ \hunit, respectively, that are also in better agreement with Planck result than with R21 result. We can see that in Taylor expansion cases, CCH data can only provide a reasonable constraint on $H_0$ in the 2-order and CCH + Pantheon + MCT data are in better agreement with Planck result than with R21 result.

When we also consider including GP off-diagonal covariance matrix (GPmat) to minimize $\chi^2$ functions, as listed in Table \ref{h0res}, we find that in contrast to the old $H_0$ constraints, the uncertainties of $H_0$ increase and except for the $H(z) + E(z) + E_0 + \mathrm{GPmat}$ case, the central values of $H_0$ also increase. Therefore, GPmat does not provide useful insights on alleviating $H_0$ tension.

To explore the effects of error models, fiducial models, and $H_0$ priors ($H^{\rm prior}_0$) on the results, we generate 15 samples of 128 simulated $H(z)$ data in each combination, and reconstruct $H_0$ values using GP method ($H^{\rm GP}_0$), which are summarized in Table \ref{summockh0}. We find that both error models and fiducial models play insignificant roles on altering the $H^{\rm GP}_0$ values, except that the former can result in different magnitudes of uncertainties. Our original choice of Error Model 1 is an intermediate one with the uncertainties neither too big as those from Error Model 4 nor too small as those from Error Model 2, thus it is a reasonable choice.

By comparing the central values of $H^{\rm GP}_0$ with $H^{\rm prior}_0$, we find that for most of the samples, $H^{\rm GP}_0\geq H^{\rm prior}_0$, where the maximum differences (Max $\Delta H_0$) between central values of $H^{\rm GP}_0$ and values of $H^{\rm prior}_0$ are less than $1\sigma$. Although the values of $H^{\rm GP}_0$ depend on $H^{\rm prior}_0$, they are consistent with each other within $1\sigma$. Nevertheless, $H^{\rm GP}_0$ values tend to increase relative to the GP reconstructed $H_0$ values from the 31 CCH data. Therefore, the influence of $H^{\rm prior}_0$ is trivial for our purpose.

\begin{figure*}
\centering
\includegraphics[width=0.8\linewidth]{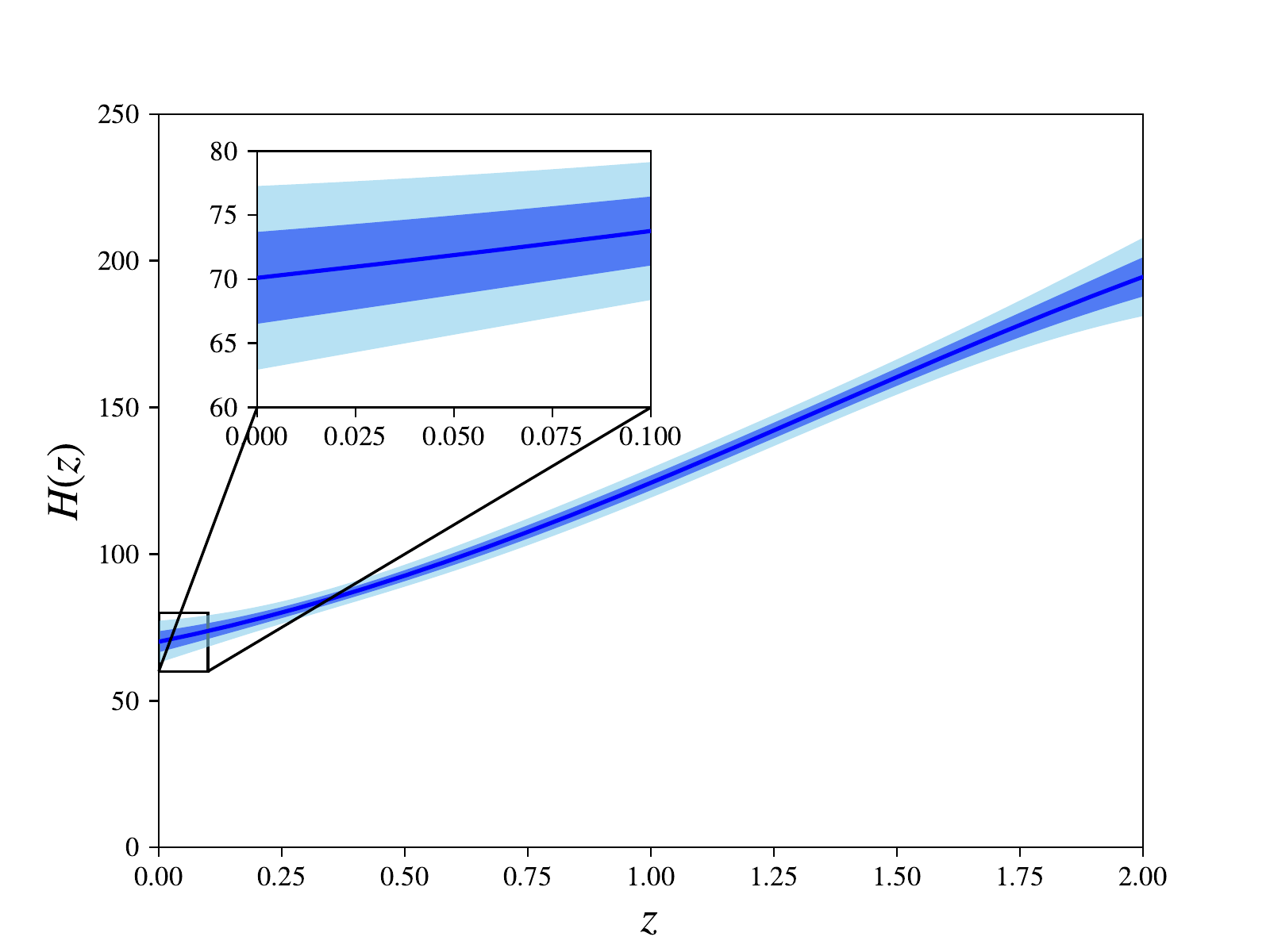}
\caption{Reconstructed $H(z)$ from simulated data generated in flat $\Lambda$CDM fiducial model by using the Gaussian Processes method. The blue solid line is the mean of the reconstruction, and the dark and light blue shaded regions are $1\sigma$ and $2\sigma$ errors, respectively.}
\label{gpmockhz}
\end{figure*}

\begin{figure*}
\centering
\includegraphics[width=0.8\linewidth]{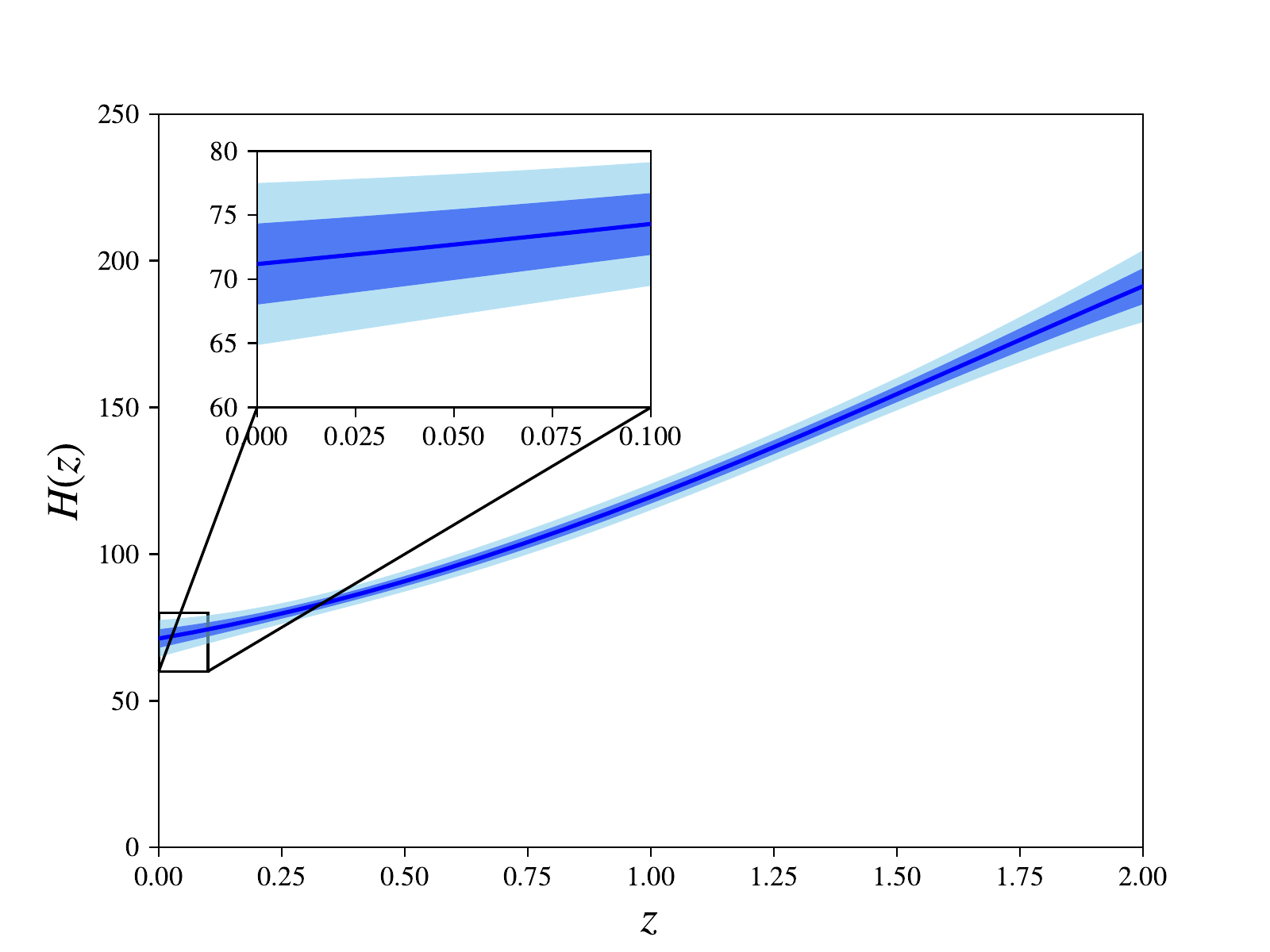}
\caption{Same as Fig. \ref{gpmockhz} but with flat CPL parametrization as fiducial model.}
\label{gpmockhzCPL}
\end{figure*}

\begin{table*}
  \centering
  \caption{$H_0$ constraints from direct Gaussian process (GP) reconstruction and $\chi^2$ (eq. \ref{chi2}) minimization with GP reconstructed expansion rate $E_{\rm GP}(z)$ (eq. \ref{Egp}).} \label{sumh0}
\begin{threeparttable}
\setlength{\tabcolsep}{5.3mm}{
 \begin{tabular}{lccccc}
 \toprule
     Method & Fiducial model & Number of $H(z)$ & $H_0$\tnote{a}  & $\Delta H_0$\tnote{b} & $\Delta H_0^{\prime}$\tnote{c}\\
 \midrule
  & -- & 19\tnote{d}  & $64.9\pm 4.2$ & $-0.59\sigma$ & $-1.89\sigma$ \\
 Gaussian Process & -- & 31 & $67.46\pm4.75$ & $0.01\sigma$ & $-1.17\sigma$ \\
 Reconstruction & Flat \lcdm\ model & 128 & $71.10\pm3.58$ & $0.75\sigma$ & $-0.81\sigma$ \\ 
  & Flat CPL parametrization & 128 & $71.18\pm3.16$ & $1.18\sigma$ & $-0.59\sigma$ \\ 
 \midrule
 & -- & 31 & $70.41\pm1.58$ & $1.82\sigma$ & $-1.36\sigma$ \\
 $\chi^2$ minimization & Flat \lcdm\ model & 128 & $72.11\pm1.43$ & $3.11\sigma$ & $-0.56\sigma$ \\ 
  & Flat CPL parametrization & 128 & $71.34\pm1.39$ & $2.67\sigma$ & $-0.98\sigma$ \\ 
 \bottomrule
 \end{tabular}}
\begin{tablenotes}[flushleft]
\item[a] \hunit.
\item[b] Differences between our results and Planck value of $H_0=67.4\pm0.5$ \hunit.
\item[c] Differences between our results and R21 value of $H_0=73.2\pm1.3$ \hunit.
\item[d] \cite{Busti:2014dua}.
\end{tablenotes}
\end{threeparttable}
\end{table*}

\begin{table*}
  \centering
  \caption{Marginalized $1\sigma$ constraints on $H_0$ from flat \lcdm\ and Taylor expansion.}  \label{taylortable}
\begin{threeparttable}
\setlength{\tabcolsep}{7.4mm}{
 \begin{tabular}{lcccc}
 \toprule
 Model & Data set & $H_0$\tnote{a}  & $\Delta H_0$\tnote{b} & $\Delta H_0^{\prime}$\tnote{c}\\
 \midrule
 Flat \lcdm & CCH & $67.77\pm 3.13$ & $0.12\sigma$ & $-1.60\sigma$ \\
 & CCH + Pantheon + MCT  & $69.19\pm 1.84$ & $0.94\sigma$ & $-1.78\sigma$ \\
 \midrule
 2-order Taylor expansion & CCH & $66.50\pm3.92$  & $-0.23\sigma$  & $-1.62\sigma$  \\
 & CCH + Pantheon + MCT & $68.45\pm1.90$ & $0.53\sigma$ &  $-2.06\sigma$ \\
 \midrule
 3-order Taylor expansion & CCH & $70.63\pm8.26$ & $0.39\sigma$  & $-0.31\sigma$ \\
 & CCH + Pantheon + MCT & $68.62\pm1.96$ & $0.60\sigma$  & $-1.95\sigma$ \\
 \midrule
 4-oder Taylor expansion & CCH & $68.00^{+8.00}_{-10.00}$ & $0.06\sigma$  & $-0.64\sigma$ \\
 & CCH + Pantheon + MCT & $68.75\pm1.97$ & $0.66\sigma$  & $-1.89\sigma$ \\
 \bottomrule
 \end{tabular}}
\begin{tablenotes}[flushleft]
\item[a] \hunit.
\item[b] Differences between our results and Planck value of $H_0=67.4\pm0.5$ \hunit.
\item[c] Differences between our results and R21 value of $H_0=73.2\pm1.3$ \hunit.
\end{tablenotes}
\end{threeparttable}
\label{modeltable}
\end{table*}

\begin{table*}
  \centering
  \caption{$H_0$ constraints from minimizing $\chi^2$ functions.}
\begin{threeparttable}
\setlength{\tabcolsep}{9.4mm}{
\begin{tabular}{cccc}
    \toprule
    data & $H_0$ (\hunit) & $\Delta H_0$ (Planck) & $\Delta H_0$ (R21) \\
    \midrule
    $H(z)+E(z)$ & $70.41\pm1.58$ & 1.82$\sigma$ & $-1.36\sigma$ \\
    $H(z)\mathrm{mat}+E(z)$\tnote{a} & $72.34_{-1.92}^{+1.90}$ & 2.49$\sigma$ & $-0.37\sigma$ \\
    $H(z)+E(z)+E_0$ & $70.13\pm1.49$ & 1.74$\sigma$ & $-1.55\sigma$ \\
    $H(z)\mathrm{mat}+E(z)+E_0$ & $71.56\pm1.79$ & 2.23$\sigma$ & $-0.87\sigma$ \\
    $H(z)+E(z)+{\rm GPmat}$\tnote{b}  & $70.81_{-3.02}^{+2.64}$ & 1.11$\sigma$ & $-0.95\sigma$ \\
    $H(z)\mathrm{mat}+E(z)+{\rm GPmat}$ & $77.45_{-3.63}^{+3.61}$ & 2.74$\sigma$ & 1.10$\sigma$ \\
    $H(z)+E(z)+E_0+{\rm GPmat}$  & $69.95_{-2.48}^{+2.27}$ & 1.01$\sigma$ & $-1.24\sigma$ \\
    $H(z)\mathrm{mat}+E(z)+E_0+{\rm GPmat}$ & $73.69\pm2.73$ & 2.27$\sigma$ & 0.16$\sigma$ \\
    \bottomrule
  \end{tabular}}
\begin{tablenotes}[flushleft]
\item[a] $H(z)$mat stands for CCH data including the ones with full covariance matrix.
\item[b] GPmat stands for GP with off-diagonal covariance matrix.
\end{tablenotes}
\end{threeparttable}
  \label{h0res}
\end{table*}

\begin{table*}
  \centering
  \caption{Average values of GP reconstructed $H_0$ (Avg $H^{\rm GP}_0$) from 128 simulated $H(z)$ data in $H^{\rm GP}_0\geq H^{\rm prior}_0$ samples of 15 randomly well-behaved samples in each Error Model (1, 2, 3, or 4) and fiducial model (\lcdm\ or CPL) with each different $H_0$ prior ($H^{\rm prior}_0=$ 67.4, 70, or 73.2 \hunit). Max $\Delta H_0$ represents the difference between the maximum $H^{\rm GP}_0$ and $H^{\rm prior}_0$. Avg $\Delta H_0$ represents the difference between Avg $H^{\rm GP}_0$ and $H^{\rm prior}_0$.}
%Error Model 1: $\sigma_+=16.25z+18.46$, $\sigma_0=11.82z+10.56$, and $\sigma_-=7.40z+2.67$; Error Model 2: $\sigma_+=22.00z+10.00$, $\sigma_0=14.75z+6.00$, and $\sigma_-=7.50z+2.00$; Error Model 3: $\sigma_+=10.89z+22.96$, $\sigma_0=11.51z+11.17$, and $\sigma_-=6.35z+8.37$; Error Model 4: $\sigma_+=8.97e^z-16.36z+18.07$, $\sigma_0=10.28e^z-16.70z+6.57$, $\sigma_-=-1.20e^z+8.21z+9.19$.}}
\begin{threeparttable}
\setlength{\tabcolsep}{2.7mm}{
  \begin{tabular}{ccccccc}
    \toprule
    Error Model & Fiducial model & $H^{\rm prior}_0$ (\hunit) & $H^{\rm GP}_0\geq H^{\rm prior}_0$ samples & Max $\Delta H_0$ & Avg $H^{\rm GP}_0$ & Avg $\Delta H_0$ \\
    \midrule
     &  & 67.4 & 10/15 & 0.35$\sigma$ & $68.21\pm3.46$ & $0.23\sigma$ \\
     & \lcdm & 70 & 10/15 & 0.36$\sigma$ & $70.74\pm3.31$ & $0.22\sigma$ \\
     \multirow{2.5}{*}{1} &  & 73.2 & 10/15  & 0.36$\sigma$ & $73.92\pm3.29$ & $0.22\sigma$ \\\cmidrule{2-7}
     &  & 67.4 & 11/15 & 0.50$\sigma$ & $68.26\pm3.03$ & $0.28\sigma$ \\
     & CPL & 70 & 11/15 & 0.54$\sigma$ & $70.84\pm3.19$ & $0.26\sigma$ \\
     &  & 73.2 & 11/15 & 0.47$\sigma$ & $73.93\pm3.10$ & $0.26\sigma$\\
    \midrule
     &  & 67.4  & 10/15 & 0.77$\sigma$ & $68.35\pm2.41$ & $0.39\sigma$ \\
     & \lcdm & 70  & 10/15 & 0.54$\sigma$ & $70.93\pm2.43$ & $0.38\sigma$\\
     \multirow{2.5}{*}{2} &  & 73.2  & 10/15 & 0.55$\sigma$ & $74.01\pm2.45$ & $0.33\sigma$ \\\cmidrule{2-7}
     &  & 67.4  & 11/15 & 0.65$\sigma$ & $68.40\pm2.27$ & $0.44\sigma$\\
     & CPL & 70  & 11/15 & 0.70$\sigma$ & $70.91\pm2.45$ & $0.37\sigma$\\
     &  & 73.2  & 11/15 & 0.70$\sigma$ & $74.13\pm2.41$ & $0.39\sigma$ \\
    \midrule
     &  & 67.4  & 10/15 & 0.66$\sigma$ & $68.28\pm3.70$ & $0.23\sigma$\\
     & \lcdm & 70  & 10/15 & 0.66$\sigma$ & $70.90\pm3.76$ & $0.24\sigma$\\
     \multirow{2.5}{*}{3} &  & 73.2  & 10/15 & 0.65$\sigma$ & $74.13\pm3.77$ & $0.25\sigma$ \\\cmidrule{2-7}
     &  & 67.4  & 10/15 & 0.40$\sigma$ & $68.13\pm3.51$ & $0.21\sigma$\\
     & CPL & 70  & 11/15 & 0.42$\sigma$ & $70.77\pm3.61$ & $0.21\sigma$\\
     &  & 73.2  & 10/15 & 0.48$\sigma$ & $74.00\pm3.58$ & $0.22\sigma$\\
    \midrule
     &  & 67.4  & 10/15 & 0.34$\sigma$ & $68.34\pm4.73$ & $0.20\sigma$\\
     & \lcdm & 70  & 10/15 & 0.35$\sigma$ & $70.95\pm4.73$ & $0.20\sigma$\\
     \multirow{2.5}{*}{4} &  & 73.2  & 10/15 & 0.31$\sigma$ & $74.08\pm4.69$ & $0.19\sigma$\\\cmidrule{2-7}
     &  & 67.4  & 10/15 & 0.48$\sigma$ & $68.61\pm4.32$ & $0.28\sigma$\\
     & CPL & 70  & 10/15 & 0.50$\sigma$ & $71.37\pm4.35$ & $0.31\sigma$\\
     &  & 73.2  & 10/15 & 0.40$\sigma$ & $74.47\pm4.24$ & $0.30\sigma$\\
    \bottomrule
  \end{tabular}}
% \begin{tablenotes}[flushleft]
% \item[a] \hunit.
% %\item[b] $H(z)$mat stands for CCH data including the ones with covariance matrix
% \end{tablenotes}
\end{threeparttable}
 \label{summockh0}
\end{table*}

\section{Conclusion}
\label{sec:conclusion}

We reconstruct $H(z)$ function from 31 CCH data by using Gaussian process (GP) and use it to combine 6 $E(z)$ compressed from Pantheon + MCT SNe Ia data to constrain $H_0$ by minimizing $\chi^2$ function (\ref{chi2}). We obtain a more restrictive value of $H_0=70.41\pm1.58$ \hunit\ that lies in the middle of the flat \lcdm\ \cite{Planck:2018vyg} TT,TE,EE+lowE+lensing $H_0$ (Planck) value and the local \cite{Riess:2021fzl} $H_0$ (R21) value, slightly closer to the latter, than the GP reconstructed value of $H_0 = 67.46 \pm 4.75$ \hunit\ \citep{Yang:2019fjt} from CCH data. When we include the full covariance matrix of 15 CCH data, we find that the GP reconstructed value of $H_0=67.06\pm4.66$ \hunit\ with lower central value and uncertainty.

Meanwhile, we forecast two sets of simulated $H(z)$ data based on CCH data using two different fiducial models and use them to predict the potentiality of future CCH data on alleviating $H_0$ tension. We find that GP reconstructed $H_0$ values from simulated data are higher and more restrictive than that from CCH data and $H_0$ constraints from $\chi^2$ minimization are closer to R21 value, which might be due to the choice of error model, fiducial models, and $H_0$ prior. When we explore the effects of error models, fiducial models, and $H_0$ priors, we find that the GP reconstructed $H_0$ results are not very sensitive to the choice of error models and fiducial models, except that different choices of the former result in different magnitudes of $H_0$ uncertainties. However, our original choice of Error Model 1 appears to be reasonable since the uncertainties are medium-sized. Although derivations of $H_0$ are dependent on the choice of $H_0$ priors in the fiducial models, the trend of GP reconstructed $H_0$ from simulations being higher than those from CCH data remains the same.

Moreover, we also use CCH and $E(z)$ data to constrain $H_0$ in the flat \lcdm\ model and cosmographical model -- Taylor expansion of Hubble parameter. We find that CCH data can only constrain 2-order Taylor expansion and favor $H_0$ values closer to Planck value than to R21 value in flat \lcdm\ and 2-order Taylor expansion. CCH and $E(z)$ data together provide more restrictive $H_0$ constraints that are closer to Planck value than to R21 value, than those from CCH data alone.

Qualitatively, we can conclude that more $H(z)$ data in the future would push $H_0$ constraints higher towards middle of Planck and R21 values and $E(z)$ data favor higher values of $H_0$. Therefore, more $H(z)$ data in the future would have the potential to alleviate $H_0$ tension and better-quality $H(z)$ data would provide even better perspective towards $H_0$ tension. Of course, one would expect different behaviors of $H(z)$ data in reality from our simulated data, since our simulations are only extrapolated from the 31 CCH data we used. However, there is no doubt that uncertainties of CCH data in the future will decrease both systematically and statistically.

\section*{Acknowledgements}
We thank Yungui Gong and Adri\`a G\'omez-Valent for helpful comments and discussions. We also thank the editor and referee for extremely useful comments which helped us improve our manuscript.

\section*{Data availability}
%%%%%%%%%%%%%%%%%%%%%%%%%%%%%%%%%%%%%%%%%%%%%%%%%%
The data underlying this article are available in the article and can be found in the papers cited in Sec. \ref{sec:data}.

\bibliographystyle{mnras}
\bibliography{myref} % if your bibtex file is called example.bib

%\begin{thebibliography}{99}
%\end{thebibliography}

%%%%%%%%%%%%%%%%%%%%%%%%%%%%%%%%%%%%%%%%%%%%%%%%%%

%%%%%%%%%%%%%%%%% APPENDICES %%%%%%%%%%%%%%%%%%%%%

%\appendix

% If you want to present additional material which would interrupt the flow of the main paper,
% it can be placed in an Appendix which appears after the list of references.

%%%%%%%%%%%%%%%%%%%%%%%%%%%%%%%%%%%%%%%%%%%%%%%%%%

% Don't change these lines
\bsp	% typesetting comment
\label{lastpage}
\end{document}